# Is Magnification Consistent?

## Why people from amateur astronomers to science's worst enemy have some basic physics wrong, and why.

Christopher M. Graney
Jefferson Community & Technical College
1000 Community College Drive
Louisville KY 40272
christopher.graney@kctcs.edu
www.jefferson.kctcs.edu/faculty/graney

*This paper is a discussion of the physics of magnification in telescopes. Special attention is given to the question of whether telescopes magnify stars. Telescopes do magnify star images, although opinions to the contrary abound.*

Is the phenomenon of magnification by a converging lens inconsistent and therefore unreliable? Can a lens magnify one part of an object but not another? Physics teachers and even students familiar with basic optics[1] would answer "no", yet many answer "yes". Many telescope users believe that magnification is not a reliable phenomenon in that it does not work for stars. This belief was central to the arguments of one of science's most prominent modern critics – a great story of how misunderstanding basic optics helped to yield bad ideas about science. So magnification is a great topic! It is accessible to students. It gives students insight into the workings of a familiar device like a telescope that even frequent telescope users often lack. And it has a fascinating side story about how misunderstanding basic science led to interesting consequences.

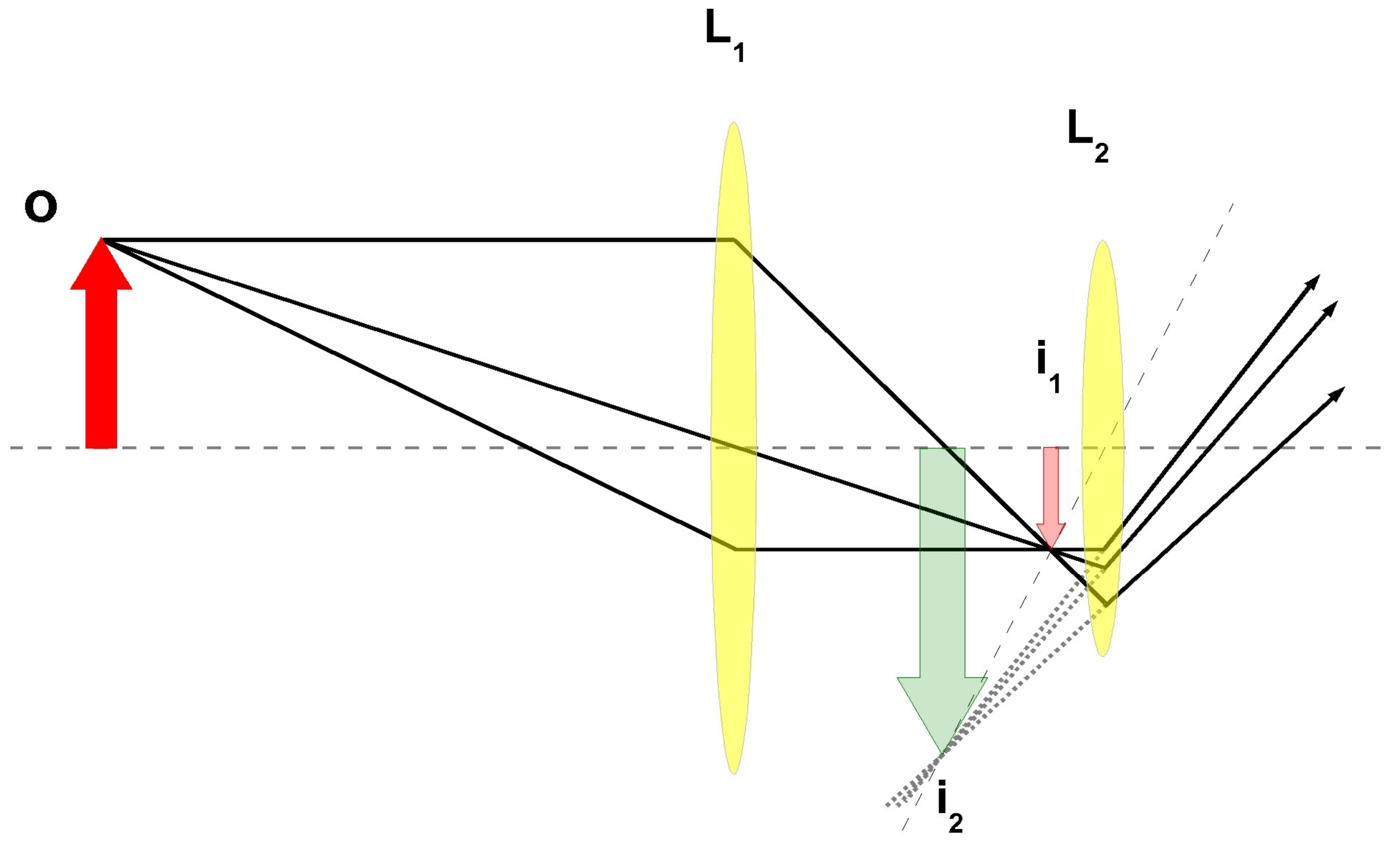

**FIGURE 1: A telescope. The "objective" lens $L_1$ forms a real image $i_1$ of the object $o$. The "eyepiece" lens $L_2$ then magnifies $i_1$, creating an enlarged virtual image at $i_2$.**

1 The optics principles and methods used in this paper use are common to many physics texts. See, for instance, Rayond A. Serway's *Principles of Physics* (Saunders College Publishing, Forth Worth, 1994), chapters 26, 28.

A simple telescope consists of a converging lens that forms a real image of a distant object and a second converging lens that magnifies that real image to produce an enlarged virtual image (Figure 1). It illustrates a basic concept in optics – that in an optical system consisting of more than one element, the image formed by the first becomes the object for the second.[2]

Let's examine the role of the second lens. Imagine using a converging lens to project a real image of a distant object onto a white screen (Figure 2). You can then use

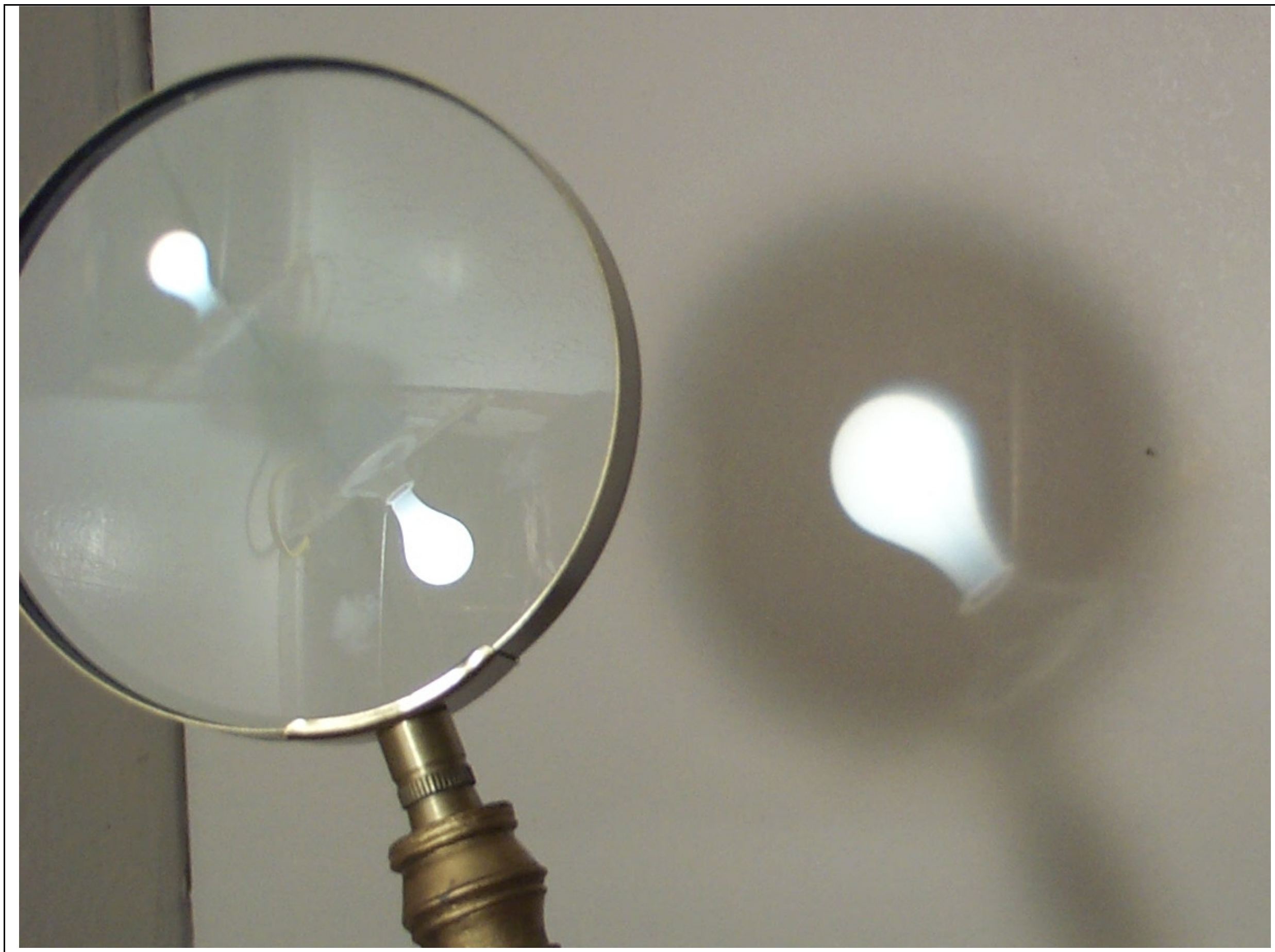

**FIGURE 2**

2 A.J. Cox, Alan J. DeWeerd, "The Image Between the Lenses: Activities with a Telescope and a Microscope," *The Physics Teacher* **41**, 176-177 (March 2003).

another magnifying (converging) lens to examine that image. A stronger magnifier enlarges the projected image more; a weaker magnifier enlarges it less. The magnifier will enlarge everything on the screen equally, be it dirt smudge or projected image. Exchanging the white screen for a translucent screen and examining the image from the rear changes nothing.[3] Everything is still enlarged equally; the real image exists and can be examined and magnified from the rear whether a screen is there or not.

Mount the two lenses in a tube and students will recognize the device as a telescope, with an "objective" lens and an "eyepiece" lens. We have established that the magnifier/eyepiece enlarges the real image equally. A telescope must equally enlarge the moon, Jupiter, and the building down the street.

However, many telescope users (amateur astronomers) hold that telescopes do not do this because they fail to enlarge stars. The conventional wisdom that "telescopes do not magnify stars" is typically explained by saying that stars are so far away that they are merely dimensionless points of light.[4] Discovery of the "special case" of stars is often attributed to Galileo Galilei.

Galilei wrote in his *Starry Messenger* of 1610 that stars "...when seen with a telescope, by no means appear to be increased in magnitude in the same proportion as other objects...." He also wrote that while planets are clearly seen to be round like the moon, the stars "...are never seen to be bounded by a circular periphery, but rather have the aspect of blazes whose rays vibrate around them and scintillate a great deal. Viewed with the telescope they appear of a shape similar to that which they present to the naked eye...."[5]

---

3 Cox and DeWeerd discuss this use of a translucent screen in their article.

4 This can be found in a variety of sources spanning a great deal of time. Examples:
"Understanding Binocular Exit Pupils" by Gary Seronik (04/18/2009), http://www.garyseronik.com/?q=node/13 (Seronik is an associate editor of *Sky & Telescope* magazine).
"OK, So What Can I Do with My New Small Telescope?" by Mike Weasner (3/26/2006), http://www.meade4m.com/articles/archive/4M_Weasner_1143408274.html (Meade is one of the most prominent manufacturers of small telescopes).
Michael Covington, *Astrophotography for the Amateur*, 2nd ed. (Cambridge University Press, 1999), pg. 170.
John Davis, *Elements of Astronomy* (J. B. Lippincott & Co., Philadelphia, 1868), pg. 164.

But physicists know this cannot be true. A star image formed by a real lens manufactured by human beings, with all its attendant imperfections, cannot be *truly* dimensionless. Moreover, it will not even be *theoretically* dimensionless, even from the perspective of basic geometric optics. Stars have a finite size. They are not *infinitely* far away. The size of an object ($s_o$), its distance from the lens ($d_o$), the size of an image ($s_i$) and the image's distance from the lens ($d_i$) are related by

$$s_i/s_o = - d_i/d_o \qquad (1)$$

Even for a star $d_o$ is not infinite, so $s_i$ will not be truly zero – a star image will have a finite size. Wave optics tells us that the image will actually be a circular aperture diffraction pattern (Figure 3) whose Airy Disk Radius $\theta_A$ depends on the telescope's aperture $D$ and the wavelength of light $\lambda$.

$$\theta_A = 1.22\lambda/D \qquad (2)$$

$\theta_A$ does not depend on the brightness of the star, but the visible size of the central maximum (and the potential visibility of the rings of the pattern) does. Imagine being in

5 Galileo Galiei, Johann Kepler, *The Sidereal Messenger of Galileo Galilei, and a part of the preface to Kepler's Dioptrics,* translation with notes by E.S. Carlos (Rivingtons, London, 1880), p. 38, p. 40. Galileo's telescope actually used a diverging lens for the eyepiece, placed ahead of where the real image formed, to create an enlarged erect virtual image.

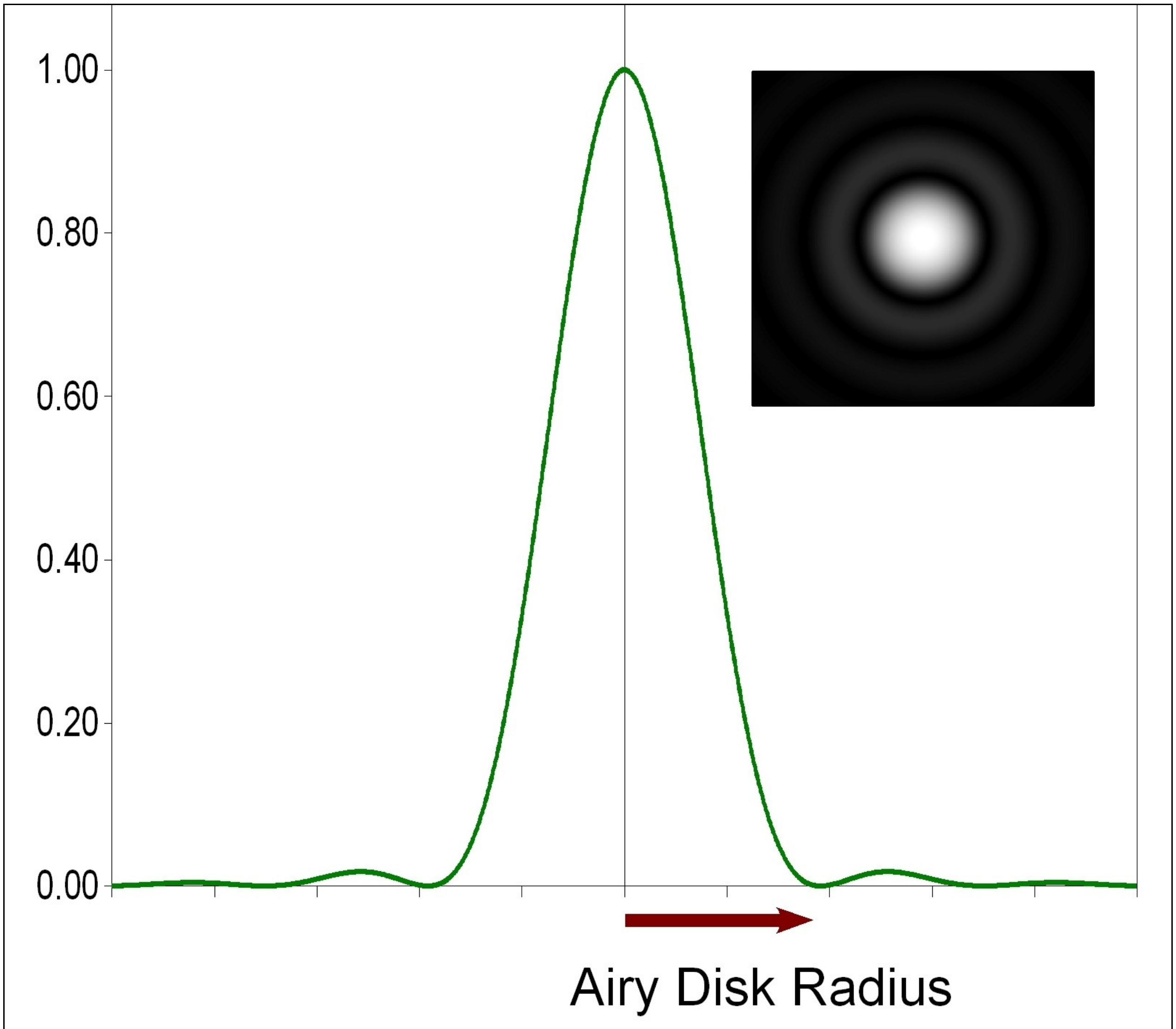

**FIGURE 3: Plot of intensity vs. radius for circular aperture diffraction pattern (insert), showing Airy Disk radius at which intensity first drops to zero. Note that the intensity is in theory only zero at the Airy Disk radius, but in the insert the first dark zone has considerable width, meaning that the size of the visible central maximum is less than the Airy Disk radius.**

a large cave made of dark rock with no light source other than a single, weak candle. The candle may "illuminate" everything in the cave, but the distant walls of the cave will appear black – the level of illumination is below what the eye can detect. Similarly, while in theory only at $\theta_A$ (and at other zero points) is a diffraction pattern truly "dark", to the eye the "dark" region is more extensive. The lower the overall intensity, the more extensive the "dark" regions are and the smaller the visible central maximum is (Figure 4). Wave optics says that star images are larger than geometric optics would indicate, and that star image size is a function of star brightness.[6]

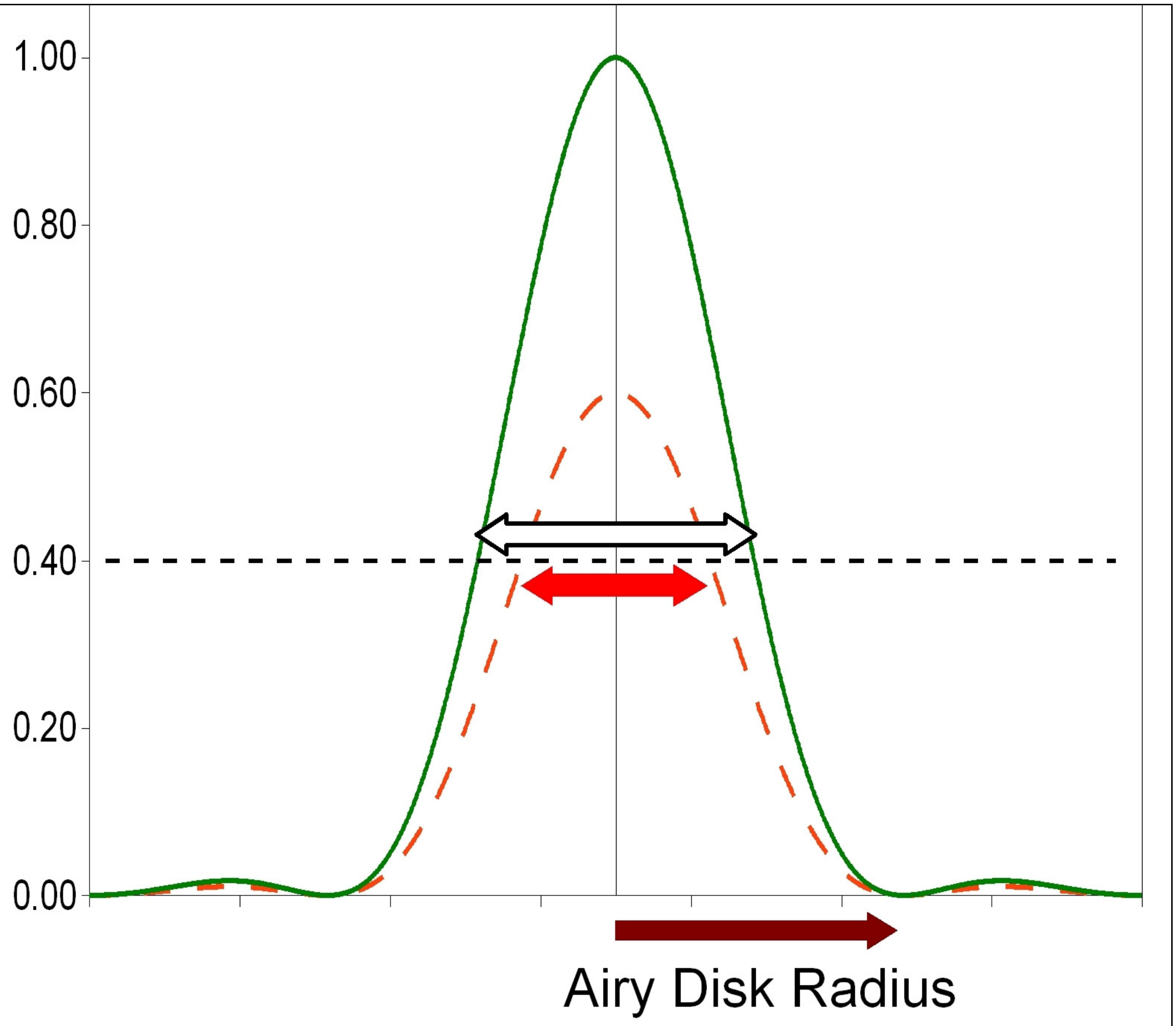

**FIGURE 4: Intensity vs. radius plots for stars of differing brightness. Both have the same Airy Disk radius but because of the eye's detection threshold (dotted line) the brighter star has a larger visible central maximum (topmost arrow) than the dimmer star (second arrow). Both stars will appear to be round with a definite size, with the brighter star appearing larger. Both will be enlarged by magnification.**

6 C. Graney, "Objects in Telescope are Further Than They Appear", *The Physics Teacher* **47**, 362-365 (September 2009).

In summary, physics says that for a telescope viewing a bright star and a faint star in the same field, the image of the bright star is larger than that of the faint star.[7] When magnified with an eyepiece, both must enlarge equally. If the moon, a planet, or the branches of a tree are also in the field of view, the star images must be magnified just as much as the other objects. Magnification in telescopes is no different than magnification elsewhere.

Then whence the conventional wisdom that telescopes do not magnify stars? The answer is that the eye cannot perceive really small objects. Imagine a bacterium that emits intense light. Looking at a glass slide that holds this luminous bacterium you see only a point of light. Even when aided by a magnifying lens your eye cannot perceive the bacterium. The lens enlarges visible features on the slide – dirt, fingerprints, the slide itself – but the point of light is unchanged. Doesn't the lens magnify everything except the light point? No. The lens enlarges everything equally; were your eyes keen enough to perceive the bacterium in the first place you would see that. With sufficient magnification (obtained via a microscope, for example) the bacterium becomes visible, and responds to further enlargement just like the other features on the slide.

7 These "images" are not the stars themselves but the diffraction patterns. [For arxiv readers – there are additional complexities involved in the magnification of star images not discussed here. As the size of the star image depends on the detection threshold of the eye, changes in that threshold should cause changes in image sizes. Significant changes in the optical system or observing conditions have been recognized as changing the image size. This was noted by William Herschel, who discusses star image sizes in his "Experiments for Ascertaining How Far Telescopes Will Enable Us to Determine Very Small Angles, and to Distinguish the Real from the Spurious Diameters of Celestial and Terrestrial Objects: With an Application of the Result of These Experiments to a Series of Observations on the Nature and Magnitude of Mr. Harding's Lately Discovered Star," *Philosophical Transactions of the Royal Society of London* **95** (1805). He describes star images of double stars as shrinking in comparison to their separations (which is sometimes misstated as shrinking in absolute terms) with significantly increased magnification (pp. 40-44). It was also noted by Christian Huygens, who observed the effect of smoked glass on star sizes in his *Systema Saturna*, (Hague, 1659) (p. 7).]

The same is true for star images. Modern telescopes have large enough apertures that the diffraction patterns that comprise stellar images are too small for the eye to detect at commonly used magnifications; the stars appear to be mere points of light. But if the images are sufficiently magnified, the diffraction patterns, or at least their central maxima, become visible. The stars appear round and respond to further magnification just like any other object. In very small telescopes such as those used by Galileo[8] an experienced observer can see this clearly even at low magnification.

But what of the *Starry Messenger*? That was merely a first impression (Galileo had only begun using telescopes in 1609). Within a few years Galileo's skill and instruments improved to the point that he indeed identified stars as being round and with measurable "disks" of definite sizes – sizes which varied with brightness. He was measuring disks by 1617[9]; in 1623 he wrote in detail on stars, saying that "every star is a perfect globe" and that their sizes showed measurable variation in that "no fixed star subtends even 5 seconds[10], many not even 4, and innumerable others not even 2"[11]; in his famous *Dialogue* of 1632 he wrote that "...the apparent diameter of a fixed star of the first magnitude is no more than 5 seconds,... one of the sixth magnitude[12] measures [5/6 seconds]...."[13] Galileo's first impression of the stars might seem to support the conventional wisdom, but his overall observations support what physics says.[14]

---

8 Aperture roughly an inch.

9 Leos Ondra, "A New View of Mizar," *Sky & Telescope* (July 2004), pp. 72-75.

10 seconds of arc (1/3600 of a degree).

11 Galileo Galilei, "Reply to Ingoli", in M. Finocchiaro, *The Galileo Affair – A Documentary History* (University of California Press, Los Angeles, 1989), p. 180 and p. 174.

12 A star of first magnitude star is bright, one of sixth magnitude is the faintest visible to an average unaided eye under dark skies.

13 Galileo Galilei, *Dialogue Concerning the Two Chief World Systems -- Ptolemaic and Copernican,* translated by S. Drake with foreword by Albert Einstein, 2nd edition (University of California Press, Los Angeles, 1967), p. 359.

14 The sizes Galileo gives are consistent with the expected diameters of visible central maxima of diffraction patterns formed by the telescopes he used. C. Graney, "On the Accuracy of Galileo's Observations," *Baltic Astronomy* **16**, 443 (2007).

But bad physics can yield good stories, and the conventional wisdom about magnification has spawned a good story. In the late 20th century the philosopher Paul Feyerabend made a name for himself with his criticism of science. He has been hailed as "the worst enemy of science".[15] His ideas still touch off controversy, as in 2008 when scholars at La Sapienza, a prestigious university in Rome, spearheaded protests over a papal visit, citing the pope's referencing Feyerabend's ideas.[16] Feyerabend used Galileo as one of his prime examples in critiquing science. Central to this was the issue of magnification. Feyerabend claimed it was not rational for Galileo to use the telescope to support the Copernican theory, because the telescope did not magnify consistently, and if the telescope is inconsistent then it is unreliable as a tool for studying the heavens. According to Feyerabend, all Galileo did was to use one idea that was weak at the time (that the telescope is a reliable) to back up another idea that was weak at the time (that the planets circle the sun).[17] To Feyerabend, Galileo's opponents had the more reasonable arguments. And yes, Feyerabend cites the *Starry Messenger*. Had Paul Feyerabend possessed a good understanding of magnification[18], science's "worst enemy" might have been a little less critical.

The question of correct understanding of the phenomenon of magnification, something that is easily accessible to students and that has practical application to the "real world" of familiar optical devices, has had a real impact beyond just the world of physicists. Misunderstanding this bit of basic science helped produce broad misunderstanding of science in general. Now there's a story to tell your students when they complain about your insisting they get the details right in your class! After all, you wouldn't want one of them to someday become the newest "worst enemy of science" based on a misunderstanding of basic physics.

---

15 John Preston, Gonzalo Munévar, David Lamb, *The Worst Enemy of Science? Essays in Memory of Paul Feyerabend* (New York, Oxford University Press, 2000).

16 "Papal visit scuppered by scholars", BBC News (January 15, 2008), http://news.bbc.co.uk/2/hi/europe/7188860.stm

17 Paul Feyerabend, *Against Method*, 3rd edition (Verso, New York, 1993, reprinted 2001). See pp. 86-105, p. 125, p. 92.

18 Or had another philosopher possessed sufficient understanding so as to correct him.